**Deterministic integration of quantum emitters and optical cavities in a van der Waals crystal**

James Liddle-Wesolowski[1,2,^], Otto Cranwell Schaeper[1,2,^], Nathan Coste[1,*], Benjamin Whitefield[1,2], Evan Williams[1], Helen Zhi Jie Zeng[1], Mehran Kianinia[1,2], Anastasiia Zalogina[1,2,*], Igor Aharonovich[1,2]

[1] School of Mathematical and Physical Sciences, University of Technology Sydney, Ultimo, New South

Wales 2007, Australia

[2] ARC Centre of Excellence for Transformative Meta-Optical Systems, University of Technology Sydney, Ultimo, New South Wales 2007, Australia

^ Contributed equally to this work

* Corresponding author: anastasiia.zalogina@uts.edu.au, nathan.coste@uts.edu.au

**Abstract**

*Single-photon emitters in hexagonal boron nitride (hBN) combine bright optical emission with optically addressable spin states, offering a promising platform for integrated quantum photonics. However, their stochastic creation and spectral variability have prevented deterministic integration with photonic cavities. Here we demonstrate a fabrication protocol that enables precise, deterministic coupling of pre-selected visible emitters to circular Bragg grating (CBG) cavities in hBN. By patterning etched alignment markers and performing pre-fabrication confocal mapping, we locate emitters with sub-micron accuracy and design cavity geometries matched to their zero-phonon line wavelengths. The resulting devices show enhanced emission and reliable spectral alignment between emitter and cavity mode. This work establishes a deterministic cavity-emitter integration scheme in a van der Waals material and provides a scalable route towards on-chip quantum photonic and spin-based platforms using hBN.*

Solid state single-photon emitters (SPE) have emerged as a versatile and scalable platform for quantum photonic technologies[1-6]. Integrating SPEs into fabricated photonic devices enables efficient coupling between the emitter and confined optical modes. This is an important prerequisite for both practical enhanced photon detection, as well as for fundamental studies of cavity quantum electrodynamics (QED), Purcell enhancement and quantum nonlinearities[7, 8]. Integrated emitter-cavity systems therefore form essential building blocks for the development of scalable quantum architectures, multiqubit networks, and photonic quantum processors[9, 10].

Amongst a variety of SPEs contenders, SPEs in hexagonal boron nitride (hBN) offer a variety of intriguing properties such as high brightness, on demand engineering, spin-photon interface and potential wavelength selectivity[11-15]. Out of the many studied defects in hBN, only two classes of emitters have been successfully integrated into photonic devices: the B centre, with zero phonon line (ZPL) at 436 nm[16, 17], and an ensemble of boron vacancy ($V_B^-$) defects[18]. Owing to their compatibility with post-fabrication activation methods, both have been incorporated into prefabricated devices[19-21]. However, while the B centre is a coherent SPE[22, 23], it does not exhibit spin - photon interface, that is highly desirable for quantum information[24].

On the other hand, the $V_B^-$ centre is an excellent spin defect with a triplet ground state but so far can not be isolated at a single level site.

Beyond these well characterised defects, there is a distinct class of visible emitters in hBN that has attracted attention for exhibiting both bright single photon emission and optically addressable spin properties[25-29]. This unique combination of optical and spin properties at a single photon level showcases compelling opportunities for quantum devices and applications. However, unlike the previously mentioned defects, there are currently no proven methods to generate the visible emitters deterministically with spatial or spectral control. While high density of SPEs can be activated by a high temperature annealing process[26], their spatial and spectral distribution remains random. This poses a significant challenge for realisation of an integrated photonic circuitry with these SPEs.

To address this challenge, we present a deterministic fabrication protocol for integrating photonic cavities with pre-selected visible emitters in hBN. Inspired by approaches used for epitaxial semiconductor quantum dots, our method combines spatially encoded reference markers and pre-fabrication confocal characterisation[30-36]. First, we identify emitters with desirable properties (i.e. high brightness, narrow ZPL, spin activity) within flat and undisturbed regions of an hBN. Once the emitter position is determined with sub-micron precision relative to the etched marker arrays, a circular Bragg grating (CBG) cavity is designed and fabricated around the emitter. This approach avoids post-fabrication defect creation and enables highly-accuracy spectral and spatial alignment, resulting in enhanced emission while preserving the optically detected magnetic resonance (ODMR) contrast.

The concept of our deterministic fabrication approach for integrating photonic structures with emitters in van der Waals materials is illustrated in Figure 1. The method relies on establishing a coordinate system defined by etched reference markers and using it to determine the precise spatial position of each emitter. As shown schematically in Figure 1a, the emitter coordinates within the hBN flake are measured optically relative to the marker array. In the next step, CBG photonic cavities are designed and fabricated by aligned electron-beam lithography, enabling high-precision placement with respect to the markers (Figure 1b). The defined pattern is then transferred into the hBN using reactive-ion etching followed by mask removal. A representative scanning electron microscope (SEM) image of a fabricated CBG cavity is shown in Figure 1c.

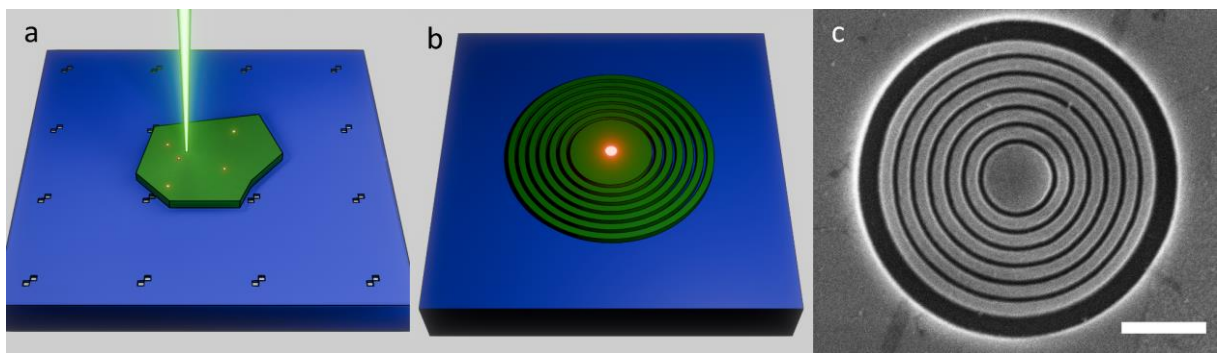

*Figure 1.* Concept of the deterministic fabrication approach. *a.* Schematic of an hBN flake on a $SiO_2$ substrate with etched alignment markers spaced 25 μm apart and the excitation laser. *b.* Schematic of a CBG structure etched into an hBN flake showing photon emission from the cavity centre. *c.* SEM image of a CBG structure etched into an hBN flake. Scale bar: 1 μm.

To ensure stable and reliable referencing throughout the fabrication process, the spatial information is encoded into a grid of alignment markers etched directly into the $SiO_2$ substrate. Alignment markers are conventionally made of metal, however, our fabrication workflow requires high-temperature oxygen annealing to activate the visible emitters in hBN. Under such conditions, metallic markers would deform, melt, or lose edge definition, resulting in blurred marker boundaries and significantly reduced alignment accuracy. To overcome this limitation, we fabricate the markers by etching them into the substrate itself using a combination of photolithography and reactive-ion etching (see methods). Etching produces thermally robust, high-contrast markers with sharp edges that remain intact throughout annealing. The markers are spaced 25 µm apart, ensuring that several markers are visible within a single 100 × 100 µm confocal scan, which enables precise coordinate extraction and reduces the impact of scan distortion.

As a light-extraction and enhancement structure, we implement CBGs, a set of concentric rings arranged to satisfy the second-order Bragg diffraction condition which enables efficient out-of-plane redirection of emission from an SPE. To design the CBGs, electric-field intensity profiles and distributions were numerically simulated using the finite-difference time-domain method implemented in Lumerical Inc. software (Figure 2). The simulated structure consisted of an hBN flake of thickness 191 nm placed on a $SiO_2$ substrate ($n_{hbn}$ =2.1, $n_{sio2}$ =1.5)[28,29]. For the base design (scaling factor of 1), the CBG comprised six concentric rings with ring widths of 117 nm, gaps of 63 nm, and a central disk of 360 nm radius.

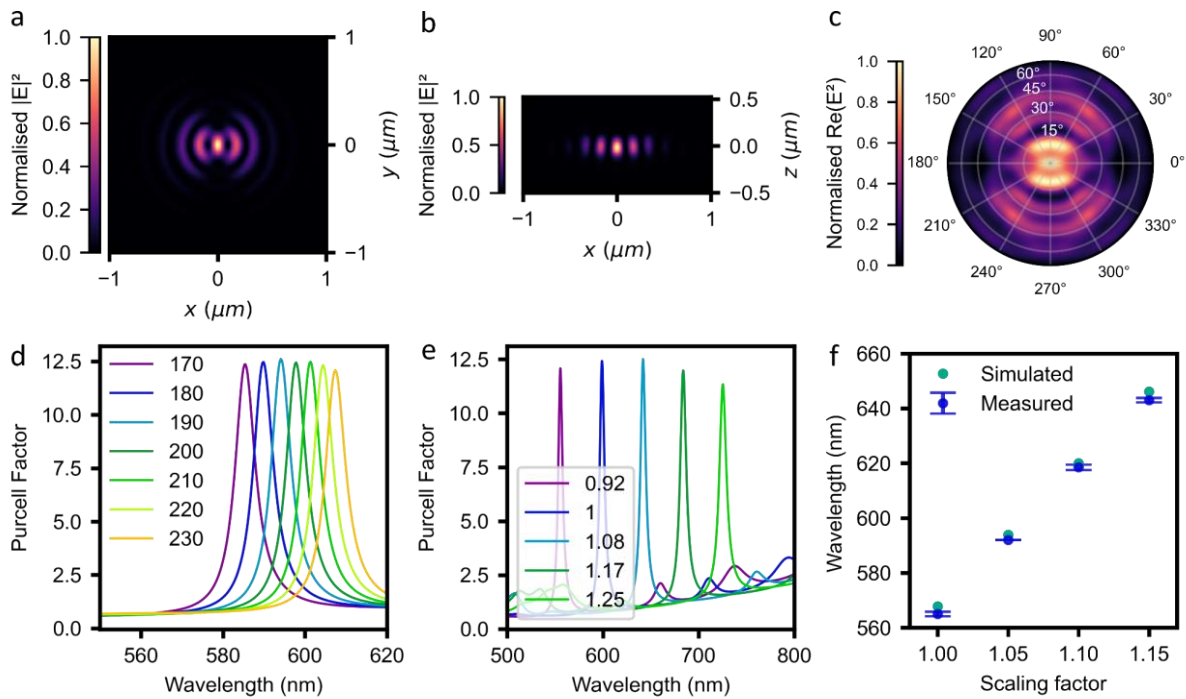

*Figure 2. CBG design. **a.** Simulated |E|² distribution in the XY plane for a flake thickness of 200 nm and a scaling factor of 1. **b.** Simulated |E|² distribution in the XZ plane under the same condition as (a). **c.** Simulated far-field emission intensity of the CBG structure under the same conditions as (a, b). **d.** Simulated Purcell factor as a function of wavelength for varying hBN thickness (values in nm). The simulated CBG has a scaling factor of 1. **e.** Simulated Purcell factor as a function of scaling factors for varying CBG wavelengths. The simulation was*

*performed with a flake thickness of 200 nm. **f.** Comparison of simulated and measured cavity mode wavelengths as a function of CBG scaling factor.*

To ensure sufficient optical confinement, both the in-plane and out-of-plane electric field distributions were simulated, as shown in Figure 2a, b. The near-field simulations confirm that the periodic CBG structure selectively reflects and confines light at the target wavelength though constructive interference governed by the Bragg phase-matching condition: $\lambda_B = \frac{2n_{eff}\Lambda}{N}$, where $\lambda_B$ - is the Bragg wavelength, which corresponds to the ZPL of the selected hBN emitter, $n_{eff}$ - is the effective refractive index of the optical mode, $\Lambda$ - is the grating period, and $N$ - is the Bragg order. The far-field emission profile (Figure 2c) confirms that the CBG efficiently directs a large fraction of the emission into a narrow ~15° cone, which is an important characteristic for high-collection efficiency quantum photonic devices.

In order to quantify the emission enhancement, the Purcell factor was calculated using an in-plane dipole source located at the centre of the disk: $F = \frac{\Gamma}{\Gamma_0}$, where $\Gamma$ and $\Gamma_0$ are the radiative decay rates with and without CBG structure, respectively. These simulations were used to establish the initial CBG geometry and guide the scaling required to match each emitter's ZPL wavelength.

To validate the robustness of the design across different cavity geometries and flake conditions, Purcell factors were also simulated for a range of hBN flake thicknesses and CBG scaling factors, as shown in Figure 2(c, d), respectively. These calibration results show that both the spectral position and magnitude of the Purcell enhancement depend on the flake thickness and cavity dimensions. In particular, thinner flakes shift the resonance to shorter wavelengths (purple line of 170 nm thickness in Figure 2c), whereas increasing the scaling factor produces a systematic red shift of the cavity mode (light green line of 1.25 scaling factor in Figure 2d). Overall, our modelling suggests Purcell factors of up to ~ 12.5 can be achieved. Combined, these simulations confirm that the CBG design can be reliably tuned to match a broad distribution of emitter ZPL wavelengths and that the enhancement mechanism remains robust across realistic fabrication variations.

To experimentally verify that the fabricated CBGs reproduced the simulated mode wavelengths, an initial array was fabricated on an hBN flake using electron-beam lithography with following reactive ion etching (see methods). Each row contained CBGs with different scaling factors, while columns provided duplicates for averaging. The cavities were measured optically with a confocal microscope (see Methods). The measured cavity modes showed excellent agreement with simulations as shown in Figure 2f, with a negligible average deviation of only ~2 nm. Such a calibration establishes a reliable offset for subsequent device fabrication and confirms the accuracy of the scaling procedure used to match the cavity mode to each emitter's ZPL.

Once established the parameter space for engineering the CBGs, we now shift our attention to integrate the emitters with the cavities. First, hBN flakes were exfoliated onto the marked substrate, allowing spatial coordinates to be referenced using the protocol described above. To increase the density of visible emitters, the hBN crystals were annealed at high temperature under a high oxygen flow in a tube furnace, a process known to introduce optically active defects into the hBN lattice (see Methods for more details). After annealing, the flakes were inspected under an optical microscope to ensure that the surface exhibited no significant damage or cracking that would impede subsequent CBG fabrication. hBN flakes with a thickness of approximately 200 nm were first pre-selected based on optical contrast. The final selected flake had a thickness of ~ 222 nm which was confirmed using atomic force microscopy.

Before fabricating the photonic devices, the spatial coordinates of the emitters within the selected hBN flake were determined relative to the etched marker array. A home-built confocal microscope was used to identify bright and spectrally stable single-photon emitters. The confocal objective has a bounded field of view of approximately 100 × 100 μm, which motivated the 20 μm spacing of the markers to ensure that several markers were visible in every scan and that reliable localisation could be achieved across the entire flake. During these pre-characterisation measurements, the sample was raster-scanned with an excitation laser of wavelength 532 nm, and the reflected light was recorded using an avalanche photodiode.

A confocal photoluminescent (PL) map of the hBN flake with the emitter's locations is shown in Figure 3a, where the location of the emitters are marked with green circles. A selection of emitter spectra is plotted in Figure 3b, demonstrating ZPL wavelengths spanning approximately 550–700 nm. Because the true geometric positions of the markers are known from the lithographic pattern, the second (reflected-laser) scan serves as a reference grid. Using this grid, spatial distortions such as scan warping or slight rotational offsets in the confocal images can be corrected algorithmically. This correction significantly reduces positional uncertainty and enables sub-micron accuracy in determining emitter coordinates.

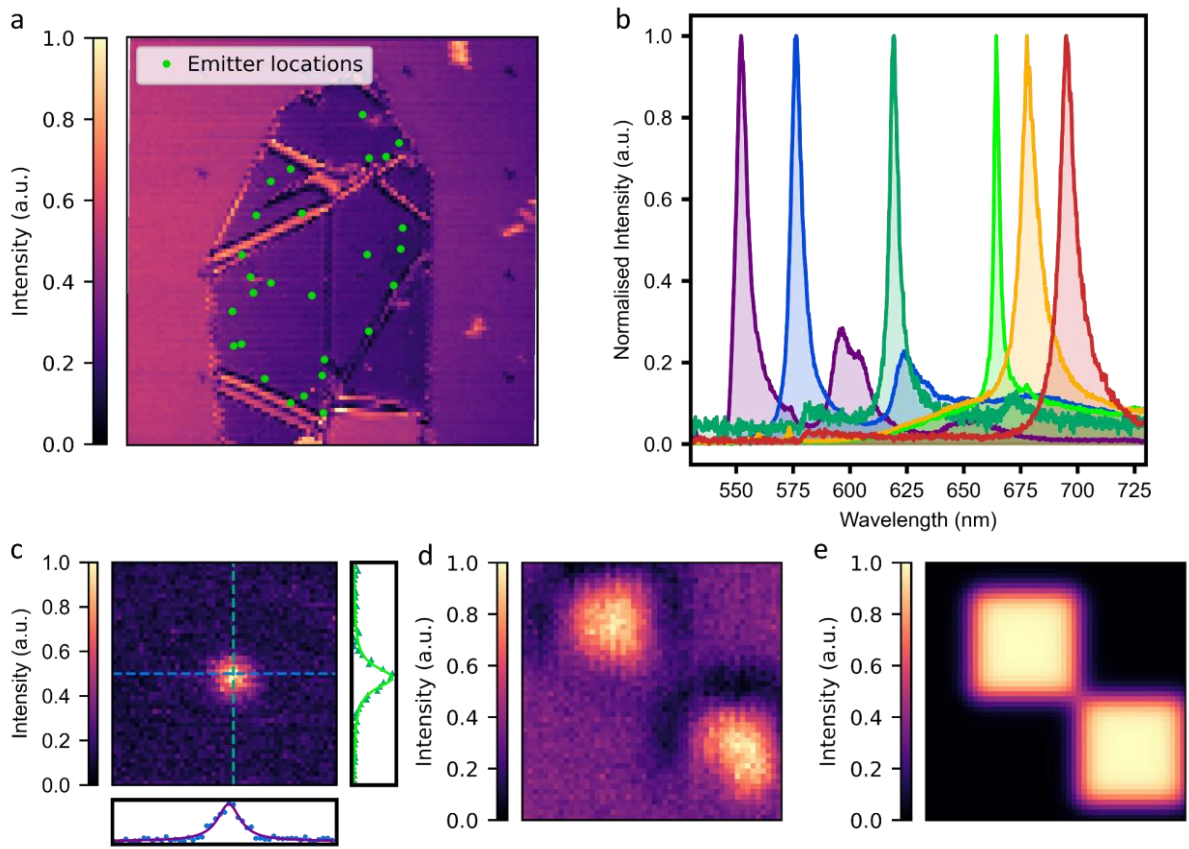

*Figure 3. Emitter characterisation before CBG structure. a. Confocal intensity map showing the excitation laser light reflected from the flake surface and alignment markers (100*

*x 100 μm). The positions of substrate markers and emitters were collected from separate scans and overlaid on the map. **b.** Representative PL spectra collected from multiple emitters across different hBN samples before structure fabrication. **c.** PL scan of an emitter (2 x 2 μm). The position uncertainty is given by the FWHM of the vertical and horizontal line profiles, corresponding to ~ 236 nm and ~ 247 nm respectively. **d.** Confocal scan of an alignment marker using the reflected laser light (2 x 2 μm). **e.** Two square fit of the marker shown in d used to determine the positional uncertainty around the marker centre, 7.4 nm (x-axis) and 6.8 nm (y-axis).*

The precise spatial coordinates were then measured on a second confocal system designed to minimise imaging distortion through the use of a closed-loop three-axis piezo stage. The sample was positioned such that several etched markers were visible within the field of view, ensuring that both emitter and marker coordinates could be referenced to the same coordinate frame (Figure 3e). To record the marker positions with high contrast, an initial scan was performed in which a controlled fraction of the excitation laser was intentionally allowed to leak through the detection path. The reflected laser light made the etched marker edges sharply visible, enabling accurate localisation. High-resolution piezo-controlled scans of individual markers were then acquired and stored together with their coordinate data.

A second set of scans was performed under standard PL conditions, with the long-pass filter inserted to suppress reflected laser light. In this configuration, only the emitters were visible. The piezo stage was used to raster across the flake and locate each pre-identified emitter. Their presence was verified by comparing their spectra to those recorded on the spectral setup. All scans, those containing marker reflections and those showing only PL from the emitters, were combined and processed using a custom distortion-correction algorithm. The known marker spacing provided fixed reference points for correcting scan warping and eliminating rotational offsets. This procedure yielded a distortion-free coordinate map in which each emitter's position was defined relative to the etched marker array with sub-micron accuracy. These corrected coordinates were then used as the input for deterministic CBG placement during electron-beam lithography.

Once the emitter coordinates relative to the markers were established, the CBG design parameters were uniformly scaled such that the cavity mode wavelength matched the measured ZPLs in Figure 3b, using the calibration procedure illustrated in Figure 2f. This ensured that each emitter would be spectrally aligned with its corresponding CBG mode. Optical image of the hBN flake with the fabricated CBG cavities is shown in Figure 4a. Figure 4b shows an SEM image of a fabricated CBG cavity around the preselected and pre-characterised emitter. A confocal map of the same CBG structure collected using reflected laser light is shown in Figure 4c.

After fabrication, the emitters positioned inside the CBG cavities were re-measured. Out of the ten integrated emitters, five exhibited clear enhancement. Representative data for one such emitter is shown in Figure 4 (d-f). Figure 4d shows the PL spectrum of the emitter before (blue curve) and after (green curve) the deterministic CBG fabrication. Among the five emitters that were studied, the ZPL intensity increased by approximately a factor of two, with the particular emitter in Figure 4b displayed a ~2.2 enhancement following CBG integration.

Figure 4e shows the saturation of the same emitter before (blue curve) and after (green curve) the incorporation with the CBG cavity. The data was fit to a three level model. Notably, enhancement of the emitter is visible with the saturation count rates at infinity increasing from ~ 81 kcounts/s to ~ 188 kcounts/s (~ 2.3 times enhancement).

Importantly, cavity fabrication did not degrade the spin properties of the emitter. The integrated emitter remained ODMR active, as shown in Figure 4f. The ODMR signature reveals both spin 1 and spin 1/2 transitions under a magnetic field of ~ 70 mT.

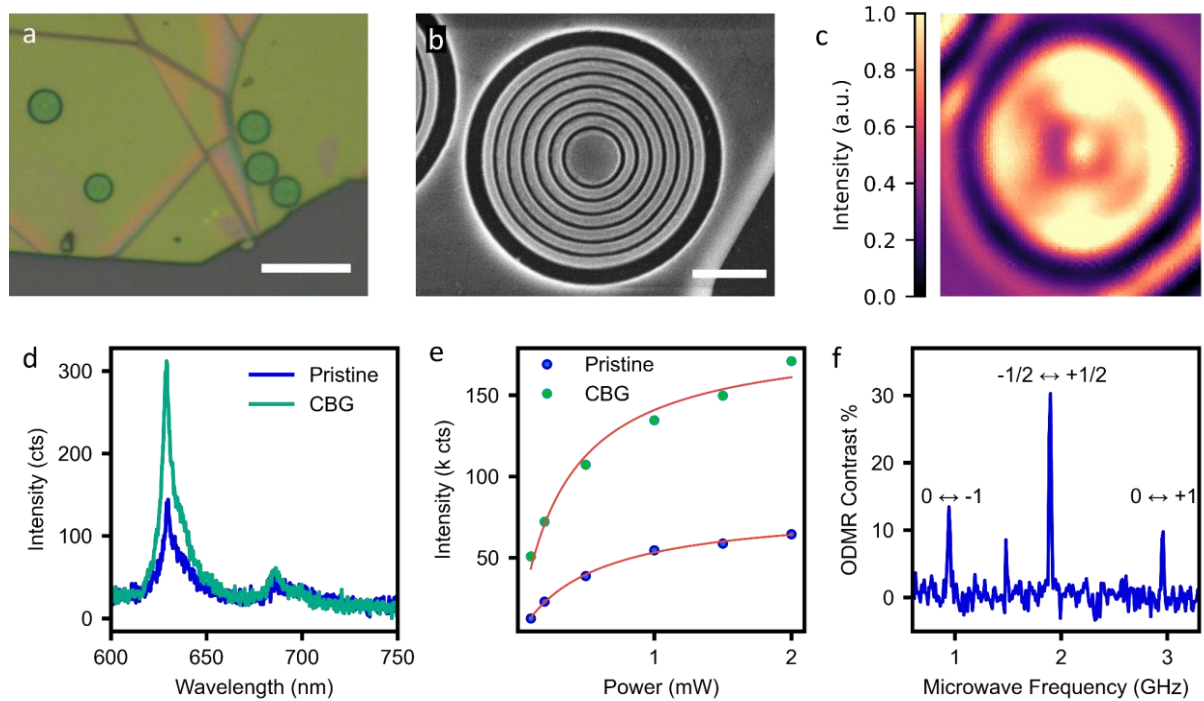

*Figure 4. CBG emission enhancement and spin characterisation. a. Optical image of the hBN flake showing the fabricated CBG structures (10 µm scale bar). b. SEM and c. confocal images of the CBG containing the emitter. (SEM scale bar corresponds to 1 µm). d. PL spectrum of an emitter on the pristine flake (blue) and the same emitter with a CBG structure fabricated around it (green). e. Saturation fits from the same emitter on the pristine flake and inside the CBG structure. f. ODMR spectrum of the same emitter showing the spin-1 and spin-1/2 under a magnetic field of ~ 70 mT. The spin-1 transitions appear around 0.95 and 2.95 GHz, the spin-1/2 transition occurs at ~1.9 GHz. The peak at ~1.48 GHz is a reflection artefact.*

**Conclusion**

We have demonstrated a fully deterministic method for integrating visible hBN emitters with circular Bragg grating cavities in a van der Waals material. By combining etched alignment markers, distortion-corrected confocal mapping, and a calibrated CBG design workflow, we achieved sub-micron emitter localisation and reliable spectral matching between each emitter's ZPL and its corresponding cavity mode. The resulting devices show enhanced PL emission, reduced saturation powers, and preserved ODMR contrast, confirming that the photonic nanostructures do not degrade the emitter's spin properties. This approach provides a scalable route to engineered light-matter interaction in hBN and establishes a foundation for integrated quantum photonic and spin-based architectures using two-dimensional materials.

**Methods**

**Device fabrication**

A silicon wafer with 285 nm of thermal oxide, was spin coated with CSAR.09 at 4000 rpm and baked on a hotplate at 180°C for 3 minutes. The wafer was then patterned using electron beam lithography (EBL) with an array of 5 by 5 mm alignment marker cells. Each alignment marker cell possessed small markers every 25 µm, with larger markers every 100 µm and 1 mm, to aid in substrate navigation during PL characterisation. The wafer was developed using CSAR developer (ARP-600-546) for 60 seconds and rinsed with IPA before blow drying with nitrogen. The alignment markers were then briefly etched by inductively coupled plasma reactive ion etching (ICP-RIE) to transfer the pattern into the SiO2 layer. Using the remaining CSAR to protect the surface, the wafer was diced into 6 by 6 mm pieces, each piece containing a 5 by 5 mm alignment marker cell. The residual CSAR was removed by placing the diced samples in N-methyl-2-pyrrolidone (NMP) at 180°C for 5 minutes, followed by rinsing in acetone and isopropanol and finally dried under $N_2$ flow.

hBN was exfoliated onto the marked substrate using the scotch-tape method. We generated emitters in the hBN flakes by annealing the sample in a tube furnace (Lindberg Blue 3000) with an oxygen flow of 1000 sccm oxygen, and heated to 1000 °C for 4 hours. The substrate was then placed into a UV-ozone chamber for 4 hours to clean the surface of the sample and remove any unstable emitters. A flake with suitable thicknesses for fabrication was first identified using optical microscopy and its thickness was measured using atomic force microscopy.

Emitters were characterised on the selected flake by use of a lab-built PL set-up, see optical characterisation section for further description. The thickness of the hBN flake and the emission wavelength of the characterised emitters were used in a finite-difference time-domain (FDTD) simulation to produce CBG designs with a mode that matched each individual emitter.

To ensure accurate spatial alignment between flakes and markers, the EBL Elionix system first referenced the global alignment markers on the substrate, utilising the in-built manual alignment procedure. Aligning the EBL pattern to the physical markers on the substrate, corrected for translation and rotation across the entire substrate. After establishing this reference frame, the emitter coordinates obtained from the confocal scans could be directly used to position the centre of each CBG with high precision, ensuring that the fabricated cavity was accurately aligned to the emitter location. The CSAR was developed in CSAR developer for 60 seconds, then 5 seconds in xylene, lastly 20 seconds in IPA, and was then dried using a N2 gun. Following development, the pattern was transferred into the hBN using an inductively coupled plasma reactive ion etcher (ICP-RIE) system using 300W RF, 60 sccm Ar, 5 sccm SF6 and 11 mTorr chamber pressure. The remaining CSAR was removed by placing the sample in 130°C CSAR remover for 1 hour, then rinsed in MiliQ water, IPA, and dried using N2 gun. Finally the emitters were re-characterised using identical confocal microscopy conditions to evaluate the effect of integration on the previously measured optical properties.

**Optical characterisation**

To obtain accurate emitter-to-marker distances, we used two complementary confocal PL microscopes, one optimised for spectral characterisation and the other for high-precision spatial mapping.

Emitter spectral characterisation and cavity mode measurements were performed using a home built confocal PL set up optimised for spectral data collection. The system used a static stage and a scanning mirror (Newport FSM-300-01). Samples were excited with a 532 nm continuous wave (CW) laser (Laser Quantum gem) though a 100x objective with an NA of 0.9 (Nikon). A 568 nm long pass filter blocked the laser emission from reaching the measuring devices. Emitter light was directed to a spectrometer (Acton SP2300i) to record spectral positions. Emitter saturation was measured by rotating a neutral density filter to change the excitation intensity while recording the corresponding photon counts with an avalanche photodiode (APD) (Excelitas).

A home built PL setup optimised for spatial data collection minimised the surface map distortion by utilising a three axis piezo stage (Physik Instrumente P-611.3). A 532 nm CW laser (CNI MGL-FN-532) through a 100x objective with an NA of 0.9 (Nikon) was used to determine precise emitter locations relative to the alignment markers. The flakes were positioned using manual stage controls to ensure multiple markers were visible along both the x and y directions. Surface scans were recorded with free space APD (Laser Components COUNT® T) without adjusting the manual sample stage. Initial surface scans had a portion of the excitation laser reflected into the detection path; this was done to enhance the contrast of the alignment markers. The Piezo stage was then centered on an alignment marker, and a high-resolution scan was saved along with the coordinate data. A second scan was performed with the 568 nm long pass filter to block the reflected laser. The piezo stage was moved to the emitter's location and the identity was confirmed by comparing the measured spectra to previously recorded data. High resolution scans and coordinate positions were recorded for each emitter.

Subsequent scans were performed to obtain complete coordinate data for all of the emitter and marker location. The recorded marker coordinates were mapped to the known real-world positions from the lithographic pattern. A custom Python-based correction algorithm determined a transformation function that described the warping and rotation between the measured and real-world marker coordinates. Applying this function to the measured emitter coordinates corrected them to their marker aligned positions. Applying this transformation to the PL map corrected it to the real-world marker positions. This approach enabled precise positioning of individual emitters relative to the alignment markers, allowing for the EBL pattern to be aligned to a sub-micron accuracy.

**ODMR measurements**

Emitter excitation was performed using a 532 nm CW laser (Laser Quantum gem) and a 100x 0.7 NA objective (Mitutoyo). Emission was collected through a 568 nm long pass filter and detected with APDs (Excelitas). Radio frequency (RF) signals were sent to the emitter via a copper wire stretched above the flake. CW ODMR measurements were performed over a frequency range set by a signal generator (AnaPico APSIN 4010). The RF signal was set with 1 ms on and 1 ms off timing and was amplified (Keylink KB0727M47C) before reaching the sample. A magnetic field was created using a NdFeB magnet (N35) that was positioned below the sample.

**Acknowledgments**

The authors acknowledge financial support from the Australian Research Council (CE200100010, FT220100053, DP250100973) the Air Force Office of Scientific Research (FA2386-25-1-4044). The authors thank the ANFF node at USYD and UTS for access to facilities.